\documentclass[aps,prl,amsmath]{revtex4} 
\usepackage{graphicx}  
\usepackage[usenames, dvipsnames]{color}
\usepackage[plainpages = false, pdfpagelabels, 
                 pdfpagelayout = useoutlines,
                 bookmarks,
                 bookmarksopen = true,
                 bookmarksnumbered = true,
                 breaklinks = true,
                 linktocpage=all,
                 pagebackref=false,
                 colorlinks = true,
                 linkcolor = BrickRed,
                 urlcolor  = blue,
                 citecolor = BrickRed,
                 anchorcolor = green,
                 hyperindex = true,
                 hyperfigures
                 ]{hyperref} 

\begin{document} 
 
\title{\href{http://necsi.edu/research/evoeco/}{Mean Field Approximation To a Spatial Host-Pathogen Model}} 
 
\author{M.A.M. de Aguiar$^{1,2}$, E. Rauch$^{1,3}$ and 
Y. Bar-\!Yam$^{1,4}$}  
 
\affiliation{$^1$ New England Complex Systems Institute, Cambridge,  
Massachusetts 02139 \\ $^2$Instituto de F\'isica Gleb Wataghin, 
Universidade Estadual de Campinas, 13083-970 Campinas, S\~ao Paulo, 
Brazil\\ $^3$MIT Artificial Intelligence Laboratory, 200 Technology 
Square, Cambridge, Massachusetts 02139 \\ $^4$Department of Molecular 
and Cellular Biology, Harvard University, Cambridge, Massachusetts 
02138}

\begin{abstract}
 
We study the mean field approximation to a simple spatial
host-pathogen model that has been shown to display interesting
evolutionary properties. We show that previous derivations of the mean
field equations for this model are actually only low-density
approximations to the true mean field limit. We derive the correct
equations and the corresponding equations including
pair-correlations. The process of invasion by a mutant
type of pathogen is also discussed. 

[This article was published as \emph{Physical Review E} \textbf{67,}
  047102 (2003).  Errata for the published version are corrected here
  and explicitly listed at the end of this document.]
\end{abstract} 
 
\maketitle 
 

Ecologists have become increasingly aware of the importance of space
in evolution and epidemiology. It has become apparent that
inhomogeneities in spatially distributed populations can fundamentally
change the dynamics of these systems
\cite{tilman,hiroki1,hiroki2,rauchprl,rauchjtb}. A simple lattice
host-pathogen model, first introduced by Tainaka \cite{tainaka}, has
become a paradigm for the study of spatially extended dynamics. In
epidemiology, the model was introduced by Comins et al. \cite{comins}
and further studied in refs. \cite{hassell,sato,rand,haraguchi}. The
model is a probabilistic cellular automaton, in which the state of
each site is updated according to the state of
nearby sites. Insight in the role of the parameters and global
behavior of the system can be obtained from the mean field
approximation, when all hosts and pathogens experience the same local
environment. This first approximation to the dynamics can be improved
by including pair correlations.
 
The mean field equations for this host-pathogen model were first 
presented by Rand et al. \cite{rand} (see also \cite{keeling1,iwasa}). 
Corrections due to pair correlations were considered in \cite{iwasa}. 
Satulovsky and Tom\'e \cite{satulovsky} have also derived the mean 
field and pair correlation equations for a similar model. In this 
paper we argue that the mean field equations and the pair 
approximation in refs. \cite{rand,keeling1,iwasa} are actually only 
approximations to the correct equations. In the derivations in these 
works, the probability of infection of a susceptible host by an 
infected individual is overcounted, as is the probability of a 
susceptible host being born on a empty site. These equations are valid 
only for small rates of transmissibility of the pathogen and for small 
birth rates of susceptible hosts, when these overcountings are not 
important. We obtain the correct mean field equations for the 
well established model of Tainaka \cite{tainaka} as well as 
the pair correlation equations. The process of invasion by a mutant
type is also discussed. 
 
 
We consider a two-dimensional spatial lattice with $N$ sites. The
state of each site can be either empty $(0)$, occupied by a
susceptible (S), or occupied by an infected individual
($I_{\tau}$). At each time step, the susceptible hosts reproduce into
each nearby cell with probability $g$ if that cell is not yet
occupied. The probability of reproduction is independent for each
neighbor. An infected host dies with probability $v$, the
virulence. Finally an infected host $I_{\tau}$ causes a neighboring
uninfected host to become infected with probability $\tau$, the
transmissibility. The subscript $\tau$ allows more than one type to be
present on the lattice. For the sake of simplicity we shall re-label
the state $(S)$ as $(1)$ and ($I_{\tau}$) as ($\tau$).
 
The state of the system is denoted by $\sigma = (\sigma_1, \sigma_2, 
...., \sigma_N)$, where $\sigma_i$ is the state at the i-th site.  We 
call $\omega_{i}(\sigma)$ the transition probability per unit time of 
the state at the site $i$. The transition probabilities are: 
\begin{equation} 
\label{prob1} 
\omega_{i}(\sigma) = \left\{  
 \begin{array}{ll} 
   1 - (1-g)^{n_i} \quad & \mbox{ if}  \quad\sigma_i = 0\\ 
   1-(1-\tau)^{m_i} \quad & \mbox{ if}  \quad\sigma_i = 1\\ 
   v \quad & \mbox{ if}  \quad \sigma_i = \tau 
 \end{array} \right.  
\end{equation} 
where $n_i=\sum_j \delta(\sigma_{i+j},1)$ is the number of susceptible
neighbors to $i$, and $m_i=\sum_j \delta(\sigma_{i+j},\tau)$ is the
number of infected neighbors to $i$. The sum over $j$ runs through all
the nearest neighbors. We call $\zeta$ the total number of nearest
neighbors. Note that, since a susceptible cannot be infected twice,
the probability of becoming infected has to be calculated as ``one
minus the probability of not becoming infected". This gives rise to
the term $1-(1-\tau)^{m_i}$.  Similarly, an empty site can become
occupied only by offspring of a single susceptible neighbor host, thus
the term $1 -(1-g)^{n_i}$.
 
Allowing for the simultaneous existence of different types of
pathogens, and mutation between the types, enables the study of
evolutionary dynamics, where different types compete for the same
susceptibles. When a pathogen of transmissibility $\tau$ reproduces,
its offspring has probability $\mu$ of having transmissibility $\tau
\pm \epsilon$.  For simplicity we assume that $\tau$ may take only
discrete values $\tau_k=k \epsilon$, $k=1,2,...,M$ where
$M=1/\epsilon$. The state occupied by a host infected with pathogen
$\tau_k$ will be labeled $(\tau_k)$. The transition probability per
unit time of the state at the site $i$ is then given by
\begin{equation} 
\label{prob} 
\omega_{ik}(\sigma) = \left\{  
 \begin{array}{ll} 
   1 - (1-g)^{n_i} \quad & \mbox{ if}  \quad\sigma_i = 0\\ 
   \Omega_k \quad & \mbox{ if}  \quad\sigma_i = 1\\ 
   v \quad & \mbox{ if}  \quad \sigma_i = \tau_{k'}           
 \end{array} \right. \ 
\end{equation} 
where $\Omega_k$ is the probability that susceptible hosts become 
infected by the pathogen with transmissibility ${\tau_k}$: 
\begin{equation} 
\label{probinf} 
\Omega_k = \chi  \left[\frac{\mu}{2} p_{k-1} + \frac{\mu}{2} p_{k+1} + 
             (1-\mu) p_k \right]  
\end{equation} 
with 
\begin{equation} 
\label{omega} 
\chi =  \frac{1-\prod_{j}(1-\tau_j)^{m_j}} {\sum_j p_j } 
\end{equation} 
and $p_k = 1-(1-\tau_k)^{m_k}$. For $\Omega_1$ and $\Omega_M$ the
terms in $p_0$ and $p_{M+1}$ should be discarded and the factor
$(1-\mu)$ replaced by $(1-\frac{\mu}{2})$.  
 
 
Approximate mean field equations for the lattice model with a single
pathogen type were obtained using simple considerations in
refs. \cite{keeling1,iwasa} in the context of the same spatial model
and in \cite{satulovsky} for a similar spatial predator-prey
model. These equations fail to take into account the fact that a
susceptible cannot be infected twice or that an empty site cannot
accommodate more than one offspring. In order to find the correct mean
field limit of the spatial model, we have derived the master equation
for the probability of the system as a whole. For the present case of
multiple pathogen types, it reads  
\begin{equation} 
\label{master2} 
\frac{\rm{d} P(\sigma,t)}{\rm{d} t} = \sum_{i=1}^N \sum_{k'} \left[  
P(\sigma^i_{k'}) \omega_{i\sigma_i}(\sigma^i_{k'}) -  
P(\sigma)  \omega_{i k'}(\sigma) \right] 
\end{equation} 
where $P(\sigma,t)$ is the probability of finding the system in the
state $\sigma$ at time $t$. The sum over $k'$ should be included only
when the argument of $\omega_{i k'}$ is $\tau_{k'}$ in the first term
and when it is $1$ in the second term. We refer to \cite{deaguiar03}
for the derivation. 
 
Given any function of the states, $f(\sigma)$, its ensemble average is 
given by $\langle f(\sigma) \rangle = \sum_{\sigma} P(\sigma,t) f(\sigma)$
Differentiating with respect to $t$ and using Eq.~(\ref{master2}) we find  
\begin{equation} 
\label{dfav2} 
 \displaystyle{\frac{\rm{d} \langle f(\sigma) \rangle}{\rm{d} t}} = 
  \sum_{i=1}^N \sum_{k'}  
  \langle [f(^i\sigma_{k'}) -f(\sigma)] \omega_{i k'}(\sigma) \rangle
\end{equation} 
where again the sum over $k'$ exists only when the argument of 
$\omega_{i k'}$ is $\tau_{k'}$ in the first term and when it is $1$ in the 
second term. \\
 

\noindent{\bf{Single type of pathogen}} -- 
In the case of a single type of pathogen the sum over $k'$ disappears 
and the transition probabilities simplify to equation 
(\ref{prob1}). To obtain an equation for the average probability of
empty sites, we consider $f(\sigma) = \delta(\sigma_i,0)$. Then 
$P_i(0,t) \equiv \langle \delta(\sigma_i,0) \rangle$ is the average 
probability that site $i$ is in the state $(0)$ in the time 
$t$. Similarly we define $P_i(1,t)$ for the average probability of 
susceptible hosts and $P_i(\tau,t)$ for the average probability of 
infected hosts.  In the approximation where the $P_i$'s are 
independent of the site, they become the mean field probabilities of 
each state, which we call $x(t)=P(1,t)$, $y(t)=P(\tau,t)$ and 
$z(t)=P(0,t)=1-x(t)-y(t)$. According to Eq.~(\ref{dfav2})
\begin{equation} 
\label{pi1a} 
\begin{array}{ll} 
 \displaystyle{\frac{\rm{d} P_i(1,t)}{\rm{d} t}} = &  
 \sum_{n=1}^N  
\langle f(^n\sigma) \omega_{n }(\sigma) -  
        f(\sigma) \omega_{n }(\sigma) \rangle 
\end{array}  
\end{equation} 
Since $f(^n\sigma)$ differs from $f(\sigma)$ only if $n=i$, only this 
term contributes  to the sum. Noticing that 
$\delta(^i\sigma,1)=\delta(\sigma_i,0)$ we get  
\begin{equation} 
\begin{array}{ll}
\label{pi1b} 
 \displaystyle{\frac{\rm{d} P_i(1,t)}{\rm{d} t}}  &=    
\langle \delta(\sigma_i,0)  \left[1-(1-g)^{n_i}\right] \rangle\\ 
 & - \, \langle \delta(\sigma_i,1)  \left[1-(1-\tau)^{m_i}\right] 
   \rangle \, .  
\end{array}
 \end{equation}

Similarly we obtain 
\begin{equation} 
\label{pi1c} 
 \displaystyle{\frac{\rm{d} P_i(\tau,t)}{\rm{d} t}}  =    
\langle \delta(\sigma_i,1)  \left[1-(1-\tau)^{m_i}\right] -  
\delta(\sigma_i,\tau)  v \rangle \, .  
 \end{equation} 

The averages can be calculated expanding the binomials $(1-g)^{n_i}$
and $(1-\tau)^{m_i}$ and approximating all pair (and higher)
correlations by simple products of one site averages
\cite{satulovsky,baalen}. We obtain 
\begin{equation} 
\label{mean1a} 
\frac{dx}{dt} = z \,  h_{\zeta}(g x) - x \,  h_{\zeta}(\tau y)
\end{equation} 
and 
\begin{equation} 
\label{mean1b} 
\frac{dy}{dt} = x \, h_{\zeta}(\tau y) - v y\;. 
 \end{equation} 
where we have defined the auxiliary function
\begin{equation} 
\label{aux} 
h_{\zeta}(\alpha) \equiv  1-(1-\alpha)^{\zeta}
 \end{equation} 

These are the correct mean field equations for the host-pathogen
model, taking fully into account the fact that a susceptible host
cannot become infected twice and that an empty site can accommodate
only one offspring. One important consequence of including this
feature, usually present in spatial models (see however
\cite{satulovsky}), is that the equations become nonlinear in $g$ and
$\tau$, losing the scaling invariance that allows one to consider only
$g + \tau+ v=1$ \cite{satulovsky}. The approximate equations 
in \cite{rand,keeling1,iwasa,satulovsky} correspond to take
$h_{\zeta}(\alpha) \approx \zeta \alpha$. \\
 
\noindent{\bf{Two types of pathogens}} -- 
When two types of pathogens are present, the competition that arises
between them gives rise to a very rich dynamics. We assume that the
two types, that we call resident and mutant, have the same virulence
$v$, but different transmissibility rates, $\tau_1$ for the resident
and $\tau_2$ for the mutant.  There are four one-site variables, $z$,
$x$, $y_1$ and $y_2$, corresponding to the probabilities of empty
sites, susceptible hosts, infected by the resident pathogen and
infected by the mutant pathogen respectively. Once again
$z=1-x-y_1-y_2$.
 
The calculation of the mean field equations in this case is more
involved, and we refer to \cite{deaguiar03} for the details. The
result is 
\begin{equation} 
\label{twom5} 
 \displaystyle{\frac{\rm{d} x}{\rm{d} t}}  =    
z \, h_{\zeta}(gx) -  
x \, h_{\zeta}(y_1\tau_1+y_2\tau_2) 
 \end{equation} 
\begin{equation} 
\label{twom9} 
\frac{dy_1}{dt} =   \bar{\chi} \, x \left\{ \frac{\mu}{2} \, 
    h_{\zeta}(\tau_2 y_2) 
    +\left(1- \frac{\mu}{2}\right) \, h_{\zeta}(\tau_1 y_1)
   \right\} - v y_1 
\end{equation} 
\begin{equation} 
\label{twom10} 
\frac{dy_2}{dt} = \bar{\chi} \, x \left\{ \frac{\mu}{2} \, 
    h_{\zeta}(\tau_1 y_1) 
    +\left(1- \frac{\mu}{2}\right) \, h_{\zeta}(\tau_2 y_2)
   \right\}   - v y_2 
\end{equation}
where
\begin{equation} 
\label{twom8} 
\bar{\chi} = \frac{h_{\zeta}(\tau_1 y_1+\tau_2 y_2)}  
{h_{\zeta}(\tau_1 y_1) + h_{\zeta}(\tau_2 y_2)} \;.
\end{equation} 

It can be shown \cite{deaguiar03} that the approximate equations lead
to complete invasion if a small amount of a more transmissible mutant
pathogen is introduced in the resident population, whereas the full
mean field equations lead to co-existence if $|\tau_2-\tau_1|$ is
sufficiently small. Invasion happens only if $|\tau_2-\tau_1|$ is
larger than a threshold that depends on $\tau_1$. 


The mean field equations can be improved by including pair 
correlations. This is done by keeping two-site probabilities 
$P_{ij}(\alpha \beta)$ in the equations while reducing higher order 
correlations to at most two-site terms. We do this reduction according 
to the truncation scheme in
\cite{baalen,mamada,kikuchi,dickman,satulovsky}. \\ 

\noindent{\bf{Single type of pathogen}} -- 
For a single type there are three possible states per site, $(0)$,
$(1)$ and $(\tau)$, and six two-site correlations. Since $\sum_j P(ij)
= P(i)$, only three of them are independent. We call the independent
correlations $u= P(10)$, $r=P(1\tau)$ and $w=P(0 \tau)$.  The other
three are given by $q \equiv P(00)=z-u-w$, $p \equiv P(11)=x-r-u$ and
$s \equiv P(\tau \tau)=y-r-w$, with $z=1-x-y$. The five independent
variables are, therefore, $x$, $y$, $u$, $r$ and $w$. The details of
the calculation can be found in \cite{deaguiar03}. The result is
  
\begin{equation} 
\label{pair4a} 
\frac{dx}{dt} = z \, h_{\zeta}(g u/z) -  x \, h_{\zeta}(\tau r/x)
 \end{equation} 
\begin{equation} 
\label{pair4b} 
\frac{dy}{dt} = x \, h_{\zeta}(\tau r/x) - v y 
 \end{equation} 
\begin{equation} 
\begin{array}{ll}
\label{pair4c} 
\displaystyle{\frac{du}{dt}} &= (q-u) \, h_{\zeta-1}(g u/z) + v r -   
  u \, h_{\zeta-1}(\tau r/x) \\
  & -\, g u [1-h_{\zeta-1}(g u/z)]
\end{array}
\end{equation} 
\begin{equation} 
\begin{array}{ll}
\label{pair4d} 
\displaystyle{\frac{dr}{dt}} &= (p-r) \, h_{\zeta-1}(\tau r/x) - v r +  
  w \, h_{\zeta-1}(g u/z) \\
  & - \, \tau r [1-h_{\zeta-1}(\tau r/x)] 
\end{array}
\end{equation} 
\begin{equation} 
\label{pair4e} 
\frac{dw}{dt} = u \, h_{\zeta-1}(\tau r/x) + v (s-w) -  
w \, h_{\zeta-1}(g u/z)  \; . 
\end{equation} 

\noindent{\bf{Two types of pathogens}} -- 
We assume once again that both the resident and the mutant pathogens
have the same virulence $v$, but different transmissibility rates:
$\tau_1$ for the resident and $\tau_2$ for the mutant. There are four
one-site variables, $z$, $x$, $y_1$ and $y_2$ and 10 two-site
variables: $u=P(10)$, $r_1=P(1\tau_1)$, $r_2=P(1\tau_2)$,
$w_1=P(0\tau_1)$, $w_2=P(0\tau_2)$, $q=P(00)$, $p=P(11)$,
$s_1=P(\tau_1 \tau_1)$, $s_2=P(\tau_2 \tau_2)$ and $s_{12}=P(\tau_1
\tau_2)$. Of these fourteen variables, only nine are independent. We
choose them to be $x, y_1, y_2, u, r_1, r_2, w_1, w_2$ and
$s_{12}$. The other five are related to them by $q=z -u-w_1-w_2$,
$p=x-u-r_1-r_2$, $s_1=y_1-w_1-r_1-s_{12}$ and $s_2=y_2-w_2-r_2-s_{12}$.
We obtain  
\begin{displaymath} 
\label{two5a} 
\frac{dx}{dt} = z \, h_{\zeta}(g u/z) -  
    x \, h_{\zeta}((\tau_1 r_1+\tau_2 r_2)/x)  
\end{displaymath} 
\begin{displaymath} 
\label{two5b} 
\frac{dy_1}{dt} =  \bar{\chi} \, x \left\{ \frac{\mu}{2} \,    
      h_{\zeta}(\tau_2 r_2/x)  
     +\left(1- \frac{\mu}{2}\right) \, h_{\zeta}(\tau_1 r_1/x)
      \right\} - v y_1
\end{displaymath} 
\begin{displaymath} 
\begin{array}{ll} 
\label{two5d} 
\displaystyle{\frac{du}{dt}} &= (q-u) \, h_{\zeta-1}(g u/z)  
       - g u [1-h_{\zeta-1}(g u/z)] \\ 
   &  - \,u \, h_{\zeta-1}((\tau_1 r_1+\tau_2 r_2)/x) + v (r_1+r_2)
\end{array}    
\end{displaymath} 
\begin{displaymath} 
\label{two5e} 
\begin{array}{ll} 
\displaystyle{\frac{dr_1}{dt}} &=  
      \bar{\bar{\chi}} \, \frac{\mu}{2} p \, h_{\zeta-1}(\tau_2 r_2/x)  
      + \bar{\bar{\chi}} \left(1-\frac{\mu}{2}\right) p \,   
      h_{\zeta-1}(\tau_1 r_1/x)  \\ 
    & + w_1 \, h_{\zeta-1}(g u/z)  
      - \tau_1 r_1 [1-h_{\zeta-1}((\tau_1 r_1+\tau_2 r_2)/x)] \\ 
    & - r_1 \, h_{\zeta-1}((\tau_1 r_1+\tau_2 r_2)/x)) - v r_1
\end{array}  
\end{displaymath} 
\begin{displaymath} 
\label{two5g} 
\begin{array}{ll} 
\displaystyle{\frac{dw_1}{dt}} & 
    = \bar{\bar{\chi}} \,  \frac{\mu}{2} u \, 
       h_{\zeta-1}(\tau_2 r_2/x)  
       + \bar{\bar{\chi}}  \left(1-\frac{\mu}{2}\right) u \, 
       h_{\zeta-1}(\tau_1 r_1/x) \\ 
    & + v (s_1+s_{12}-w_1) - w_1 \, h_{\zeta-1}(g u/z)  
\end{array} 
\end{displaymath} 
\begin{displaymath} 
\label{two5i} 
\begin{array}{ll} 
\displaystyle{\frac{ds_{12}}{dt}} &  
    =  \bar{\bar{\chi}} \frac{\mu}{2} r_2 \left\{  
         \tau_2  [1-h_{\zeta-1}(\tau_2 r_2/x)]   +  
         \, h_{\zeta-1}(\tau_2 r_2/x) \right\} \\ 
   &   + \, \bar{\bar{\chi}} \frac{\mu}{2} r_1 \left\{  
         \tau_1  [1-h_{\zeta-1}(\tau_1 r_1/x)]   +  
         \, h_{\zeta-1}(\tau_1 r_1/x) \right\} \\ 
   &    + \, \bar{\bar{\chi}} \left( 1 - \frac{\mu}{2} \right) r_2 \, 
         h_{\zeta-1}(\tau_1 r_1/x) \\ 
   &     +\,  \bar{\bar{\chi}} \left( 1 - \frac{\mu}{2} \right) r_1 \, 
         h_{\zeta-1}(\tau_2 r_2/x)  - 2 v s_{12} 
\end{array} 
\end{displaymath} 
where 
\begin{displaymath} 
\label{twom8a} 
\bar{\bar{\chi}} \equiv  \frac{h_{\zeta-1}(\tau_1 y_1+\tau_2 y_2)}  
{h_{\zeta-1}(\tau_1 y_1) + h_{\zeta-1}(\tau_2 y_2)} \;. 
\end{displaymath} 
The equations for $y_2$, $r_2$ and $w_2$ can be obtained by exchanging
the sub-indexes $1$ and $2$ in the equations for $y_1$, $r_1$ and $w_1$
respectively. 

When $\tau_1$, $\tau_2$ and $g$ are small, the functions
$h_{\zeta}(\alpha)$ can again be approximated by $\zeta \alpha$ and
the approximate pair correlation equations are obtained.  We found
that the approximate equations present limit cycles in a much larger
range of parameters than the true mean field equations
\cite{deaguiar03}. Once again, when the process of invasion is
studied, we find co-existence of similar types if $|\tau_2-\tau_1|$ is
small. However, the more transmissible pathogen type still wins over
any less transmissible ones, in the sense that either the less
transmissible type goes extinct, or its average number is always
smaller than the mutant invader. The emergence of an
intermediate-transmissibility evolutionarily stable type
\cite{rauchprl,rauchjtb} is not observed even in the pair
approximation.

\begin{figure} 
\includegraphics[height=6cm]{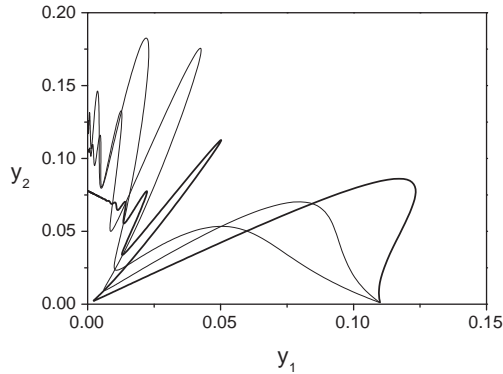} 
\caption{Invasion in the pair approximation for $g=0.05$, $v=0.2$ and 
$\tau_2=\tau_1+0.05$. The curves show the time evolution of $y_1$ and
$y_2$ (dimensionless units) for $\tau_1=0.2$ (thin), $0.3$ (thicker)
and $0.5$ (thickest). The initial conditions are $x=0.28$, $y_1=0.11$
and $y_2=0.001$. The larger the value of $\tau_1$, the closer to
extinction the population gets. } 
\label{fig1} 
\end{figure} 

However, the oscillatory approach to equilibrium revealed by the pair
approximation does give us a clue to understand how the evolutionarily
stable type appears in the spatial model. In Fig.~\ref{fig1} we show
$y_1$ versus $y_2$ for $g=0.05$, $v=0.2$ and $\tau_2=\tau_1+0.05$ for
$\tau_1=0.2$, $0.3$ and $0.5$. The initial population consists of an
equilibrium between health individuals and individuals infected by the
resident type plus a very small amount of individuals infected by the
mutant type. Although invasion occurs in all cases ($y_1$ goes to
zero), the higher the value of $\tau_1$, the closer the population of
infected hosts gets to extinction (when $y_1$ and $y_2$ get close to
zero simultaneously). If we assume that the initial population, is
structured in patches, those patches receiving the mutant type are
indeed likely to go extinct. If the patches are very large, Fig. 1
shows that the number of infected rises again after the near
extinction leading to invasion by the more transmissible
type. However, if the patches are finite, the population of infected
may die. We can estimate the minimum size of these patches so that
extinction can be prevented. If $n_p$ is the total number of sites in
the patch and $y_{1min}$, $y_{2min}$ are the values assumed by $y_1$
and $y_2$ at the near extinction time, then the actual number of
individuals (sites) infected by the resident and the mutant pathogen
types at this time is $n_p \, y_{1min}$ and $n_p \, y_{2min}$
respectively. When this number goes below $1$ there is less than one
infected site in the whole patch, and the corresponding pathogen goes
extinct. For $\tau_1=0.2$ the resident type disappears if $n_p$ is
less than $100$, whereas the mutant type disappears only if the patch
falls below $45$ sites. Typical patches observed in numerical
simulations are larger than this, implying that invasion is indeed
expected. For $\tau_1=0.5$ extinction of both resident and mutant
pathogens is prevented only if patches are larger than about $450$.
However, for $\tau_1=0.7$, the mutant pathogen with $\tau_2=0.75$, is
more likely to go extinct than the resident. If patches are larger
than about $780$ the resident pathogen survives, whereas the mutant
type disappears unless patches are larger than $890$.  Therefore, if
the actual size of the system, or patch where the mutant first appear,
is sufficiently small, extinction does happen for pathogens of large
transmissibility, stopping invasion and leading naturally to the
survival of an intermediate type.\\

\noindent Acknowledgements 
 
\noindent This work was supported in part by the National Science
Foundation under Grant No. 0083885. M.A.M.A. acknowledges financial
support from the Brazilian agencies FAPESP and CNPq.  
 

\pagebreak

\begin{center}
{\bf\large\href{http://necsi.edu/research/evoeco/}{Errata: ``Mean field
    approximation to a spatial host-pathogen model''}}
\bigskip

M.A.M. de Aguiar$^{1,2}$, E.\ Rauch$^{1,3\dag}$, B.C. Stacey$^{1,4}$ and Y. Bar-\!Yam$^{1}$

\emph{$^1$ New England Complex Systems Institute, Cambridge,  
Massachusetts 02139}

\emph{$^2$ Instituto de F\'isica Gleb Wataghin, Universidade Estadual
  de Campinas, 13083-970 Campinas, S\~ao Paulo, Brazil}

\emph{$^3$ MIT Artificial Intelligence Laboratory, Cambridge,
  Massachusetts 02139}

\emph{$^4$ Martin A.\ Fisher School of Physics, Brandeis University, Waltham, Massachusetts 02453}

\emph{$^\dag$ deceased}
\end{center}

The article ``Mean field approximation to a spatial-host pathogen
model'' by de Aguiar \emph{et al.\!}~\cite{Rauch2003} contains five
mistyped equations.  The errors were introduced in paper production
and do not affect the paper's results or conclusions.  Eq.~(\ref{pair4b})
should read
\begin{displaymath}
\frac{dy}{dt} = xh_\zeta(\tau r/x) - vy.
\end{displaymath}
As published, Eq.~(\ref{pair4b}) has an extraneous ``exponent''
$\zeta$, so that the first term reads $xh_\zeta(\tau r/x)^\zeta$.
Note that if we approximate $r = xy$ we must recover the mean-field
equation for~$dy/dt$, Eq.~(\ref{mean1b}), which has no such exponent.

In the paragraph following Eq.~(\ref{pair4e}), the variable $p$ is
defined as $P(1q)$.  This should instead read $p = P(11)$.  There are
also sign errors in Eq.~(\ref{pair4d}) and in the unnumbered equation
for $dr_1/dt$ in the pair-approximation treatment of the system with
two types of pathogen.  In each case, the sign of the third term
should be positive rather than negative.  Also, $\bar{\omega}$ should
be replaced with $\bar{\bar{\chi}}$ wherever it appears.  The correct
equations are
\begin{align}
\frac{dr}{dt} = &\ (p-r)h_{\zeta-1}(\tau r/x)
 - vr \nonumber\\
 & + wh_{\zeta-1}(gu/z)
   -\tau r\left[1 - h_{\zeta-1}(\tau r/x)\right],\nonumber
\end{align}
\begin{align}
\frac{dr_1}{dt} = &\ \bar{\bar{\chi}} \frac{\mu}{2} 
                     ph_{\zeta-1}(\tau_2r_2/x) \nonumber\\
 & + \bar{\bar{\chi}}\left(1 - \frac{\mu}{2}\right)
   ph_{\zeta-1}(\tau_1r_1/x) \nonumber\\
 & + w_1h_{\zeta-1}(gu/z) \nonumber\\
 & - \tau_1r_1\left[1-h_{\zeta-1}
                          \left((\tau_1r_1 + \tau_2r_2)/x\right)
                          \right] \nonumber\\
 & - r_1h_{\zeta-1}\left((\tau_1r_1 + \tau_2r_2)/x\right)
  - vr_1, \nonumber\\
\frac{dw_1}{dt} = &\ \bar{\bar{\chi}} \frac{\mu}{2} 
                     uh_{\zeta-1}(\tau_2r_2/x) \nonumber\\
 & + \bar{\bar{\chi}}\left(1 - \frac{\mu}{2}\right)
   uh_{\zeta-1}(\tau_1r_1/x) \nonumber\\
 & - w_1h_{\zeta-1}(gu/z) \nonumber\\
 & + v (s_1+s_{12}-w_1). \nonumber
\end{align}

These equations were replicated in an expanded treatment by de Aguiar
\emph{et al.\!}~\cite{deaguiar2004}.  The computer code used to
generate the figures in both articles used the correct equations.
Consequently, the misprints reported in this note do not affect the
results or conclusions of either paper.

\end{document}